\documentclass[sigconf]{acmart}

\AtBeginDocument{%
  }

\acmConference[FSE Companion '24]{}{June 23-27, 2025}{Trondheim, Norway}

\renewcommand\footnotetextcopyrightpermission[1]{} 
\usepackage[]{collab}
\usepackage{enumitem}
\usepackage{balance}
\usepackage{xcolor}
\usepackage{subfigure}
\usepackage{multirow}
\usepackage{multicol}
\usepackage{tabularx}
\usepackage{siunitx}
\usepackage{float}
\usepackage{color}
\usepackage{hyperref}
\usepackage{diagbox}
\usepackage{caption}
\usepackage{makecell}
\usepackage{colortbl}
\usepackage{tcolorbox}
\usepackage{mdframed}
\usepackage{listings}
\usepackage{booktabs}
\usepackage[utf8]{inputenc}
\usepackage{calligra}
\usepackage[normalem]{ulem}
\usepackage{indentfirst}
\usepackage{url}
\usepackage{soul}
\usepackage{nicematrix}
\usepackage{fontawesome5}
\usepackage{textcomp}
\usepackage{stfloats}
\usepackage{verbatim}
\usepackage{graphicx}
\usepackage{amsmath,amsfonts}
\usepackage{algorithmic}
\usepackage{array}
\usepackage[utf8]{inputenc}

\definecolor{ballblue}{rgb}{0.13, 0.67, 0.8}
\definecolor{grey}{rgb}{0.9, 0.9, 0.9}
\definecolor{googlered}{rgb}{0.914, 0.262, 0.207}

\collabAuthor{jj}{blue}{Junjie}
\collabAuthor{gb}{magenta}{Guangba}
\collabAuthor{zh}{purple}{Zhihan Jiang}
\collabAuthor{jz}{magenta}{Jiazhen Gu}
\collabAuthor{zb}{teal}{Zhuangbin}
\collabAuthor{jc}{googlered}{JC}
\collabAuthor{ry}{orange}{RY}

\newcommand\nm{L4\xspace}
\newcommand\company{\textit{Company-X}\xspace}
\newcommand\platform{\textit{Platform-X}\xspace}
\newcommand\failurenum{428\xspace}

\newcommand{\ie}{{\em i.e.},\xspace}
\newcommand{\eg}{{\em e.g.},\xspace}

\newcommand{\fixedwidth}[1]{{\ttfamily \small #1}}

\newcommand{\boxmargin}{1mm}

\newtcolorbox{myboxa}[2][]{
    colback=gray!10!white,
    colframe=black, enhanced,
    attach boxed title to top left={yshift=-2mm,xshift=5mm},
    title=#2,#1
}
\newtcolorbox{myboxb}[2][]{
    boxsep=3pt,
    left = \boxmargin, right = \boxmargin, top = \boxmargin, bottom = \boxmargin,
    title={#2},#1
}
\newtcolorbox{myboxc}{
    colback=gray!15!white,
    arc = 0pt, outer arc = 0pt,
    boxsep=0pt, left = 3pt, right = 0pt, top = 0pt, bottom = 0pt, 
    leftrule=3pt, bottomrule=0pt,toprule=0pt, rightrule=0pt,
    left = \boxmargin, right = \boxmargin, top = \boxmargin, bottom = \boxmargin
}
\newtcolorbox{myboxd}{
    colback=gray!10,
    colframe=black,
    width=\columnwidth,
    arc=1mm, auto outer arc,
    boxrule=0.5pt,
}

\definecolor{myyellow}{HTML}{FFF2CC}
\newcounter{finding}
\newcommand{\finding}[1]{\refstepcounter{finding}
 	\vspace{1mm}
	\begin{mdframed}[linecolor=gray!25,roundcorner=12pt,backgroundcolor=myyellow!30,linewidth=3pt,innerleftmargin=2pt, leftmargin=0cm,rightmargin=0cm,topline=false,bottomline=false,rightline = false]
		\textbf{Finding \arabic{finding}:} #1
	\end{mdframed}
}

\begin{document}
\title{L4: Diagnosing Large-scale LLM Training Failures via Automated Log Analysis}


\author{Zhihan Jiang}
\affiliation{%
  \institution{The Chinese University of Hong Kong}
  \city{Hong Kong SAR}
  \country{China}}

\author{Junjie Huang, Guangba Yu}
\affiliation{%
  \institution{The Chinese University of Hong Kong}
  \city{Hong Kong SAR}
  \country{China}}
\authornote{Guangba Yu is the corresponding author.}


\author{Zhuangbin Chen}
\affiliation{%
  \institution{Sun Yat-sen University}
  \city{Zhuhai}
  \country{China}}

\author{Yichen Li, Renyi Zhong}
\affiliation{%
  \institution{The Chinese University of Hong Kong}
  \city{Hong Kong SAR}
  \country{China}}


\author{Cong Feng, Yongqiang Yang, Zengyin Yang}
\affiliation{%
  \institution{Huawei Cloud}
  \city{Shenzhen}
  \country{China}}


\author{Michael R. Lyu}
\affiliation{%
  \institution{The Chinese University of Hong Kong}
  \city{Hong Kong SAR}
  \country{China}}

\renewcommand{\shortauthors}{Jiang et al.}


\begin{abstract}
As Large Language Models (LLMs) show their capabilities across various applications, training customized LLMs has become essential for modern enterprises.
However, due to the complexity of LLM training, which requires massive computational resources and extensive training time, failures are inevitable during the training process.
These failures result in considerable waste of resource and time, highlighting the critical need for effective and efficient failure diagnosis to reduce the cost of LLM training.

In this paper, we present the first empirical study on the failure reports of \failurenum LLM training failures in our production \platform between May 2023 and April 2024.
Our study reveals that hardware and user faults are the predominant root causes, and current diagnosis processes rely heavily on training logs.
Unfortunately, existing log-based diagnostic methods fall short in handling LLM training logs.
Considering the unique features of LLM training, we identify three distinct patterns of LLM training logs: cross-job, spatial, and temporal patterns.
We then introduce our \underline{L}og-based \underline{L}arge-scale \underline{LL}M training failure diagnosis framework, \nm, which can automatically extract failure-indicating information (\ie log events, nodes, stages, and iterations) from extensive training logs, thereby reducing manual effort and facilitating failure recovery.
Experimental results on real-world datasets show that \nm outperforms existing approaches in identifying failure-indicating logs and localizing faulty nodes.
Furthermore, \nm has been applied in \platform and demonstrated its effectiveness in enabling accurate and efficient failure diagnosis.

\end{abstract}

\maketitle

\section{introduction}

Large language models (LLMs) have revolutionized various fields including natural language processing~\cite{he2023large,yang2024harnessing} and software engineering~\cite{gao2023makes,li2025coca}, enabling breakthrough applications such as code generation~\cite{guo2024deepseek}, document translation~\cite{wang2023document}, and dialogue systems~\cite{yi2024survey}.
The superior performance of LLMs is primarily driven by the scaling law~\cite{kaplan2020scaling}, which establishes that the model capacity strongly correlates with both the model size and the volume of training data.
For instance, recent models like Grok-1~\cite{grok} incorporate 314 billion parameters, while training datasets such as RedPajama~\cite{weber2024redpajama} have reached 30 trillion tokens.

To achieve state-of-the-art model capability, significant efforts have been devoted to training or fine-tuning LLMs, which requires substantial computational resources.
For example, the Llama3-405B model was trained using 16,384 H100 GPUs for 54 days~\cite{dubey2024llama}.
To facilitate these demanding training requirements, IT enterprises have developed multi-tenant LLM development platforms, such as Amazon SageMaker~\cite{amazon-ai} and Google Vertex AI~\cite{googlevertex}.
These platforms allow users to submit LLM training jobs with customized hardware resources and access to specialized software libraries and tools.

LLM training failures have become the norm rather than the exception~\cite{zhang2023opt,wang2023gemini,zhong2023swift,wu2023transom,gupta2024just}, primarily due to three key factors: the immense scale and complexity of computational resources, the substantial volume of training data, and the extended duration of training processes.
For instance, during the training of Llama3-405B, Meta utilized 16,384 H100 GPUs and encountered 466 failures over a 54-day period~\cite{dubey2024llama}.
These failures result in significant losses in both computational resources and time, requiring substantial human effort for diagnosis and resolution~\cite{gupta2024just,jiang2024megascale}.
A notable example comes from BigScience's training of the BLOOM-176B model using 384 GPUs~\cite{bloom_training}.
During this procedure, each hardware failure resulted in an average loss of 1.5 hours of training time, with the recovery process consuming an additional 5 to 10 hours~\cite{bloom_training,hu2024characterization}.

\begin{figure}[t]
    \centering
    \includegraphics[width=\columnwidth]{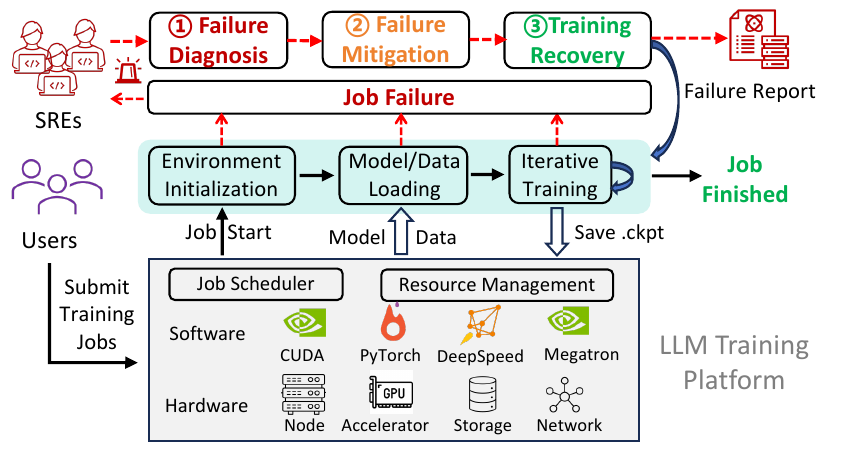}
    \caption{LLM training and failure management process in LLM development platforms.}
    \vspace{-20pt}
    \label{fig:platform}
\end{figure}

As shown in Fig.~\ref{fig:platform}, failure diagnosis is a critical step in the recovery process following a training failure.
Rapid and accurate diagnosis allows engineers to identify root causes, implement remediation strategies, and swiftly resume model training.
However, diagnosing failures in large-scale LLM training remains a time-consuming and labor-intensive task, primarily due to the challenges posed by both node-level and cluster-level complexities.
(1) \textbf{Node-level Complexity}: An AI node typically comprises several layers~\cite{yu2024surveyfailure}, including AI accelerators (\eg GPUs and NPUs~\cite{liao2021ascendnpu}), AI toolkits (\eg CUDA~\cite{cuda}), AI frameworks (\eg PyTorch~\cite{pytorch} and MindSpore~\cite{mindspore}), and AI algorithms (\eg Transformers~\cite{transformer}).
The intricate dependencies and interactions between these layers often result in a multitude of noisy fault manifestations, complicating accurate fault localization due to fault propagation.
(2) \textbf{Cluster-level Complexity}: Training large-scale LLMs training often involves thousands of AI nodes, utilizing diverse communication paradigms such as Data Parallelism (DP), Pipeline Parallelism (PP), and Tensor Parallelism (TP)~\cite{shoeybi2019megatron}. These complex structures make it challenging to quickly pinpoint faulty nodes within the vast network of interconnected components.
Therefore, it is imperative to comprehensively characterize LLM training failures and explore automation opportunities to diagnose these failures.

To facilitate this need, we present an empirical study on LLM training failures and their diagnostic procedures.
Our analysis examines \failurenum failure reports collected between May 2023 and April 2024 from \platform, a large-scale production AI platform operated by \company, a world-leading cloud vendor.
The studied LLM training jobs involve models of considerable scale, with an average size of \emph{72.8 billion parameters}, and require extensive computational resources, utilizing an average of \emph{941 accelerators} per job.
Through our study, we have obtained several valuable findings that can benefit future research on ensuring LLM training reliability.
The main findings are as follows: 
\begin{enumerate}[leftmargin=*, topsep=1pt]
    \item \textbf{Failure timing}: The majority (74.1\%) of failures occur during iterative model training, indicating that this core training process is prone to failures, often resulting in wasted training time and computational resources (§~\ref{sec:RQ1}).
    \item  \textbf{Failure causes}: While the root causes are diverse, the primary culprits are hardware and user-side faults. Notably, hardware faults  are more prevalent in LLM training compared to traditional deep learning or data processing workloads~\cite{li2013characteristic,zhou2015empirical,gao2023empirical}, highlighting the unique challenges of large-scale LLM training failure diagnosis (§~\ref{sec:RQ2}).
    \item  \textbf{Diagnosis methods}: Training logs play a critical role in diagnosing failures, with 89.9\% of cases requiring detailed manual log analysis for resolution. This underscores the importance of comprehensive log analysis in LLM training (§~\ref{sec:RQ3}).
\end{enumerate}

Although Finding 3 emphasizes the importance of logs in diagnosis, an LLM training job can produce an enormous volume of raw logs (\eg serveral TBs per day), due to the extensive number of nodes and components involved~\cite{hu2024characterization}.
Within this vast amount of log data, only a small subset of logs provides actionable insights for diagnosing failures and improving the resolution efficiency, which we refer to as \emph{failure-indicating logs}.
Unfortunately, manual identification of failure-indicating logs akin to finding a needle in a haystack.
While many studies~\cite{zhang2021onion,lin2016logcluster, lin2020fast, rosenberg2020spectrum} have focused on detecting anomalous logs in traditional software systems, we found that existing methods struggle to accurately identify failure-indicating logs in LLM training scenarios.
This limitation stems from their reliance on conventional indicators such as logging level~\cite{xu2009largescale}, event frequency~\cite{lin2016logcluster}, and error semantic~\cite{liu2023scalable}.
These traditional indicators often prove inapplicable for LLM failure (details in §~\ref{sec:limitation}).

To address the limitations of existing approaches, we introduce \nm, a \underline{L}og-based \underline{L}arge-scale \underline{LL}M training failure diagnosis framework designed to automatically identify failure-indicating information and enhance diagnostic efficiency.
\nm is designed based on three distinct patterns observed in LLM training logs, \ie \textit{cross-job, spatial and temporal patterns}.
Initially, \nm parses raw training logs into structured logs and performs cross-job filtering to eliminate noisy logs unrelated to failures.
Following this, \nm leverages the spatial and temporal patterns of logs to pinpoint failure-indicating information.
In the spatial dimension, \nm embeds parsed logs from each node into log event vectors and detects potential failure-indicating nodes and log events.
In the temporal dimension, \nm profiles the training stage of logs and discovers distinctive log sequences to localize iterations where faults occur.
Finally, these identified failure-indicting log events, nodes, stages and iterations enable engineers to efficiently and precisely understand and diagnose training failures. 
Furthermore, \nm allows engineers to summarize and confirm fault patterns based on this mined information, which are then archived in the fault library to match future similar failures.

We evaluated and deployed \nm on \platform.
Evaluation using real-world large-scale training log datasets shows that \nm achieves high accuracy in identifying failure-indicating logs (87.3\% F1-score) and detecting faulty nodes (80\% top-5 accuracy).
These results surpasses all compared approaches, with a large improvements ranging from 50.7\% to 66.6\% for log identification and 18.5\% to 43.1\% for node detection.
In addition, \nm has been successfully applied in \platform since June 2024, where it has demonstrated effectiveness in facilitating the diagnosis of LLM training failures.

The main contributions of this paper are as follows:
\begin{itemize}[leftmargin=*, topsep=0pt]
    \item We present an empirical study on large-scale distributed LLM training failures, which offers valuable findings that can benefit future research on ensuring LLM training reliability (§~\ref{sec:study}).  
    \item We introduce our deployed log-based large-scale LLM training failure diagnosis framework, \nm, which automatically extracts failure-indicating information (\ie log events, nodes, stages, and iterations) from extensive training logs, thereby facilitating efficient and effective failure diagnosis (§~\ref{sec:framework}).
    \item We evaluate \nm using real-world datasets from production LLM training jobs, demonstrating that \nm outperforms other state-of-the-art baselines.
    We also share our experience from over six months of industrial application of \nm on \platform (§~\ref{sec:eval}).
\end{itemize}

\vspace{-6pt}
\section{Background}

\subsection{LLM Development \platform}
\label{sec:background-training_platform}
\platform is a multi-tenant LLM development platform at our \company, supporting LLM training jobs for hundreds of internal users and partner companies. The platform processes hundreds of LLM training jobs daily, leveraging comprehensive hardware and software infrastructure. Specifically, \platform is equipped with substantial computing resources, including heterogeneous accelerators (\eg GPUs and NPUs), distributed storage systems, and high-performance networks (\eg RDMA over Converged Ethernet and InfiniBand).  Besides, \platform provides comprehensive software support for LLM training, incorporating commonly used architecture (\eg Ascend CANN~\cite{cann} and NVIDIA CUDA~\cite{cuda}), popular training frameworks (\eg Meagtron-LM~\cite{megatron} and DeepSpeed~\cite{deepspeed}), and essential libraries (\eg Pytorch~\cite{pytorch} and Transformer~\cite{transformer}. 

The LLM training job submission and execution workflow of \platform closely resembles that of public platforms such as Amazon SageMaker~\cite{amazon-ai} and Google Vertex AI~\cite{googlevertex}.
As depicted in Fig.~\ref{fig:platform}, when users submit an LLM training job, it first allocates the necessary resources (\eg nodes and storage) and initializes the training environment based on user requirements (\eg container images and dependent libraries).
After environment initialization, datasets and models are loaded from remote storage, and the iterative training process begins (\eg fetching data, forward passing, computing loss, back-propagation, communication and saving checkpoints). 

\begin{figure}[t]
    \centering
    \includegraphics[width=0.8\columnwidth]{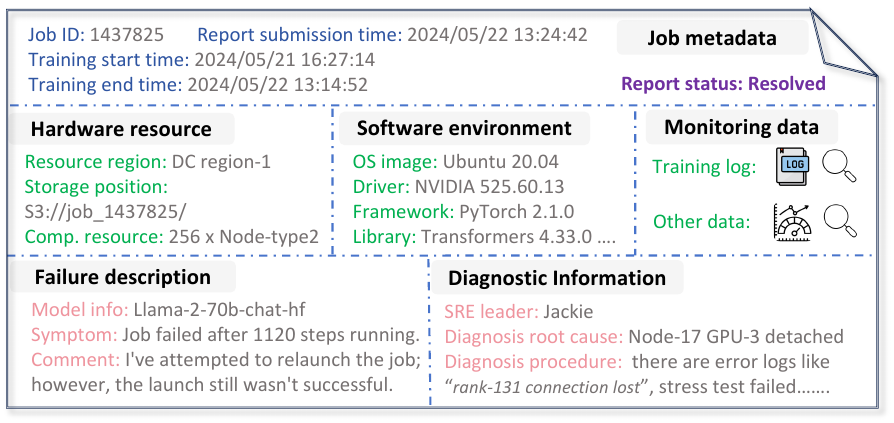}
    \vspace{-10pt}
    \caption{An example of failure reports in \platform.}
    \label{fig:failure_report}
    \vspace{-20pt}
\end{figure}

\vspace{-6pt}
\subsection{LLM Training Failure Management}

Failures are frequent and can occur at any stage during the lifecycle of an LLM training job~\cite{wang2023gemini,hu2024characterization}.
When encountering a job failure, users can submit a failure report to the failure management system in \platform to seek assistance from site
reliability engineers (SREs). 
Fig.~\ref{fig:failure_report} illustrated a failure report example in \platform, which mainly comprises five fields: \emph{job metadata}, \emph{hardware resource}, \emph{software environment}, \emph{failure description}, \emph{monitoring data} and \emph{diagnostic information}.
Each field includes several detailed sub-fields to provide comprehensive descriptions of the failed training job.
Particularly, \emph{monitoring data} are uploaded by users when they seek diagnosis help.
Training logs are one of the most commonly used monitoring data types, enabling SREs to gain an in-depth understanding of the job's status. Additionally, if users have enabled additional monitors (\eg performance and network monitors), the recorded data can also be uploaded to aid diagnosis.

Once failure reports are submitted, they are automatically assigned to appropriate SREs for handling. The SREs carefully examine the failure reports and begin the fault diagnosis process, typically involving manual inspection of monitoring data (\eg training logs). These diagnostic processes are complex and time-consuming, requiring SREs to communicate with users and other teams to identify the root cause and provide recommended solutions. After receiving feedback, users execute the suggested fixes and restart the training job. Upon successful resolution, SREs add the fault diagnostic process and root causes to the corresponding failure report and archive it within the management system, building a knowledge base for more efficient diagnosis of recurring failures.

\vspace{-5pt}
\section{LLM Training Failure Study}
\label{sec:study}

To better characterize and understand LLM training failures and their diagnosis procedures, we conduct the first empirical study on these failures in \platform.
To ensure generalizability, we avoid drawing conclusions that are specific or ambiguous.
In Sec.~\ref{sec:generalizability}, we provide a detailed discussion of the generalizability of our findings.

\vspace{-10pt}
\subsection{Study Design}

\noindent
\textbf{Study Subject.}
We collect and study \failurenum failure reports of failed LLM training jobs in \platform from May 2023 to April 2024, after eliminating duplicated reports.
These LLM training jobs encompass a diverse range of trained models (\eg LLaMA~\cite{dubey2024llama} and Vicuna~\cite{chiang2023vicuna} series), training frameworks (\eg PyTorch~\cite{pytorch} and Transformer~\cite{transformer}), and underlying hardware. 
Furthermore, all jobs in our study were all large-scale, characterized by substantial model sizes and significant computational resource usage.
Specifically, the average model size is 72.8B parameters, and the average number of accelerators utilized per job is 941.

\noindent
\textbf{Study Method.}
In this study, we comprehensively analyze all \failurenum LLM training failures and their diagnostic procedures by addressing the following research questions (RQs):
\begin{itemize} [leftmargin=10pt, topsep=2pt]
    \item \textbf{RQ1:} What are the common symptoms of LLM training failures?
    \item \textbf{RQ2:} What are the common root causes of LLM training failures?
    \item \textbf{RQ3:} What monitor data sources are typically used to diagnose LLM training failures?
\end{itemize}

We developed a taxonomy for each RQ and categorized each failure report.
To avoid potential bias, a team of five experienced SREs and Ph.D. students conduct the classification process.
Each annotator independently labeled the categories for three factors of each failure by thoroughly reviewing the documented diagnostic information.
We used Cohen’s kappa~\cite{cohen1968weighted} to assess inter-annotator agreement, achieving near-perfect agreement for each taxonomy, with all scores exceeding 0.95.
For cases with discrepancies, annotators engaged in discussions and, when necessary, consulted the submitters and corresponding SREs.
Ultimately, consensus was reached for the categorization of all \failurenum failure reports.

\vspace{-10pt}
\subsection{RQ1: Failure Symptoms}
\label{sec:RQ1}

\begin{figure}[t]
    \centering
    \includegraphics[width=0.6\columnwidth]{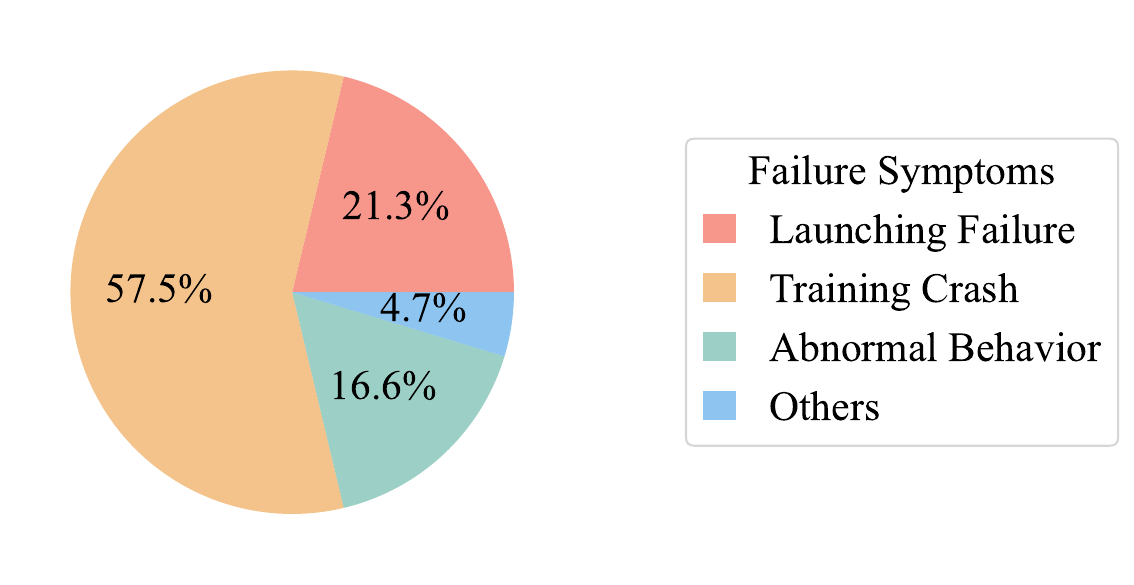}
    \vspace{-10pt}
    \caption{Classification of LLM training failure symptoms.}
    \label{fig:failure_symptoms}
    \vspace{-20pt}
\end{figure}

We first study common symptoms of LLM training failures.
A symptom is the subjective manifestation of a failure observed by users, which is manually categorized by engineers.
These symptoms are divided into four categories: \emph{launching failure}, \emph{training crash}, \emph{abnormal behavior} and \emph{others}.
The distribution of these failure symptoms across all \failurenum LLM training failures is illustrated in Fig.~\ref{fig:failure_symptoms}.

Failures occurring prior to the iterative training stage, such as during environment initialization, are classified as \emph{launching failures}.
These failures account for 21.3\% of all reported issues, as shown in Fig.\ref{fig:failure_symptoms}.
For instance, mismatches between the GPU driver and CUDA toolkit versions can cause failures during environment initialization, and misconfigurations in model parallelism can lead to errors during model loading.
The complexity and large scale of LLM training jobs pose great challenges for users and engineers in preventing these launching failures.

After the iterative training starts, the job may crash due to various reasons such as hardware faults.
These issues are frequent in LLM training due to its strong synchronization properties~\cite{wu2023transom,hu2024characterization,wang2023gemini}, \ie a local fault in a specific component (\eg a GPU or network router), can disrupt the entire training job.
According to Fig.~\ref{fig:failure_symptoms}, these \emph{training crash} failures represent 57.5\% of all failures.
Such crashes often result in waste of training time and computational resources, as they typically occur after prolonged training periods~\cite{zhang2023opt,hu2024characterization,jiang2024megascale}.
Even with the checkpointing mechanisms~\cite{eisenman2022check,zhong2023swift,wang2023gemini,gupta2024just}, the time required for failure recovery remains substantial~\cite{mohan2021checkfreq,gupta2024just}.

Moreover, certain training jobs may exhibit \emph{abnormal behaviors} like hanging or slowing down~\cite{jiang2024megascale,kokolis2024revisiting}.
For instance, an epoch might take twice as long, or training could stall at a specific iteration with no RDMA network traffic.
In such cases, users can submit failure reports to SREs for helping diagnose these issues.
Notably, such abnormal behaviors account for 16.6\% of the total failure reports.

The final category is \emph{Others}, which includes failures not directly related to a specific LLM training job, such as unavailability of platform and remote storage.
This type of failure accounts for only 4.7\% of the total failures.

\finding{Most LLM training failures (74.1\%) occur during the iterative training stage, which can waste significant computational resources and training time.}

\vspace{-5pt}
\subsection{RQ2: Failure Root Causes}
\label{sec:RQ2}

In this section, we study the common root causes of all \failurenum LLM training failures and their manifestations in monitor data, categorizing them into four categories: \emph{hardware fault}, \emph{user fault}, \emph{platform fault} and \emph{framework fault}.
Due to privacy concerns, specific distribution proportions are withheld.

\vspace{-5pt}
\subsubsection{Hardware Fault}
\label{sec:RQ2-hardware}

Similar to other LLM platforms, \platform is built with heterogeneous hardware, including nodes (\eg physical servers and virtual machines), accelerators (\eg GPUs, TPUs and NPUs), networks (\eg RoCE and InfiniBand), and remote distributed storage.
As LLM training jobs scale increase, the required hardware resources also grow, raising the probability of hardware faults~\cite{zhang2023opt,wang2023gemini,jiang2024megascale}.
Due to the synchronous properties of LLM training, a single-point hardware fault can cause the training failure, making hardware fault the most common failure root cause in our study.
This proportion is much higher than that reported in previous studies on data processing and deep learning failures~\cite{li2013characteristic, zhou2015empirical,liu2023prism,gao2023empirical,huang2024faultprofit}, indicating that LLM training procedures are more susceptible to hardware faults.

We identified four primary sub-types of hardware faults:
\begin{itemize}[leftmargin=0pt, topsep=0pt, label={}]
\item
    \textbf{Network Fault.} Training LLMs demands extensive computational resources, typically involving tens to thousands of compute nodes interconnected via high-speed networks like RoCE. The training process utilizes various parallelism paradigms, which necessitate communication between compute nodes during each iteration. As a result, network issues can impact the training process, potentially causing performance degradation, hang and failures. Therefore, among different types of hardware faults, network faults are the most common cause of training failures. When such faults occur, error log messages such as \fixedwidth{``NIC port link down''} and \fixedwidth{``increased pcs\_err\_cnt’’}~\cite{linkdown} may indicate network port failures.
\item
    \textbf{Accelerator Fault.}
    Accelerators, including GPUs, TPUs, and NPUs, are the primary computing devices for LLM training. Although the fault probability for a single accelerator is low, the overall fault probability during the training procedure is high due to the large number of accelerators involved~\cite{wu2023transom,wang2023gemini}. Similar to traditional GPU systems~\cite{tiwari2015understanding,he2023understanding}, accelerators can experience memory faults such as error correcting code (ECC) errors and stuck-at errors caused by circuit malfunctions. Power faults can also render accelerators unavailable.  Common error log messages like \fixedwidth{``double bit ecc error}'' indicating ECC memory errors~\cite{ecc} and \fixedwidth{``Aicore kernel execute failed}'' signifying computational faults~\cite{aicore}.
\item 
    \textbf{Node Fault.}
    A node (\eg a virtual machine) is an allocated unit for training jobs, containing CPUs, memory, and other resources. In large-scale clusters, node faults such as mainboard damage, power leakage, and disk errors are inevitable and can cause training failures in \platform. When a node fails, it typically becomes inaccessible, making it impossible to retrieve logs for direct fault diagnosis. In these cases, \platform relies on heartbeat mechanisms to detect node failures. The absence of regular heartbeat signals from a node is a key indicator of a node fault. 
\item 
    \textbf{Storage Fault.} The datasets, models, and checkpoints used in LLM training can be extremely large, often exceeding hundreds of gigabytes~\cite{zhang2023opt}. Users typically apply for remote distributed storage and store their data there. All nodes load data from the remote storage to start training, and checkpoints are periodically generated and stored there during the training process. Hence, any faults in remote storage can cause training failures at different stages. For example, an error log message \fixedwidth{``Failed to load checkpoint}''~\cite{checkpointstorage} may indicate issues with accessing stored model states.
\end{itemize}

\finding{LLM training procedures are vulnerable to hardware faults due to the extensive computing resources required. These faults can occur at network, accelerator, node, and storage, with network and accelerator faults being the most prevalent.}

\vspace{-10pt}
\subsubsection{User Fault}

Before submitting a failure report, users typically review their operations to attempt to resolve the issue themselves.
Despite these efforts, \emph{user fault} remains the second largest root cause of failures among all four categories, due to the complexity of user-side settings for LLM training jobs, including configurations, code, scripts, and more. Specifically, we identified four major sub-types of user faults in our study:
\begin{itemize}[leftmargin=0pt, topsep=0pt, label={}]
\item
    \textbf{Configuration Error.}
    Some LLM training failures are caused by misconfigurations in system environments and frameworks.
    When submitting LLM training jobs, users must manually configure a series of configurations.
    Even a minor misconfiguration can lead to training failures. For example, a user mistakenly set a low timeout threshold for Notify register, resulting in a timeout log message \fixedwidth{``The wait execution of the Notify register times out.''}  and subsequent training process failure~\cite{timeout}.  
\item
    \textbf{Program/Script Bug.}
    Similar to traditional software, buggy code can exist in LLM training programs and scripts, as comprehensively studied in previous empirical research~\cite{liu2019bugs,zhang2020empirical,wang2022characterizing}. 
    For instance, using inappropriate sub-process creation during training can cause the training process to get stuck~\cite{fork}.
    Since LLM training programs and scripts are typically adopted and modified from existing projects, most bugs occur in the modified parts, caused by inconsistencies between the original and modified code, such as error paths, null references and inconsistent model parameters. 
\item
    \textbf{Software Incompatibility.}
    LLM training requires specific software such as operating system images, drivers, training frameworks, libraries, and toolkits, specified by users before submitting training jobs.
    Version incompatibility is common due to independent component development~\cite{hu2024characterization,gao2023empirical}, with even minor mismatches potentially causing build or compilation failures.
    Typically, such failures could be reflected in the logs with an inappropriate version (\eg \fixedwidth{``Stream mode cannot be set in current driver version''}~\cite{version})
    Consequently, users need to carefully verify the compatibility of the relevant software versions or utilize pre-configured version information when submitting their LLM training tasks.   
\item
    \textbf{Misoperation.}
    While \platform simplifies the LLM training process, users still need to learn the operational procedures.
    Hence, users' misoperations can also result in LLM training failures.
    For example, using external remote storage for checkpoints without configuring proper access permissions can cause checkpoint writing to fail, resulting in a training crash.  
\end{itemize}

\finding{User faults constitute the predominant cause of LLM training failures due to the complexity of the settings. These faults include configuration error, program/script bug, software incompatibility and misoperation.}

\vspace{-5pt}
\subsubsection{Framework Fault}

\platform supports various open-source LLM training frameworks and libraries, including widely used options such as PyTorch~\cite{pytorch} and DeepSpeed~\cite{deepspeed}, as well as customized frameworks like CNTK~\cite{seide2016cntk}.
Like other software systems, these training frameworks are susceptible to various bugs.
Consequently, framework faults account for a small proportion of the \failurenum failures we studied, often arising from buggy code and inconsistencies during software iterations.
We have identified that, compared to widely used LLM training frameworks like PyTorch, customized LLM training frameworks are more prone to faults due to their relative immaturity.
These framework faults are particularly challenging to diagnose because they require a deep understanding of the specific training framework and significant expert effort to locate the buggy code and logic.
Moreover, fixing these bugs often requires version updates, so temporary mitigation strategies, such as version rollback, are commonly adopted until the bugs are fixed. 

\vspace{-5pt}
\subsubsection{Platform Fault}

\platform is a large-scale, multi-tenant platform that provides comprehensive support for LLM training. 
Despite careful design and iterative updates, platform-side faults are inevitable, causing the least proportion of LLM training failures.
These faults arise from various system defects, with the most common type involving resource management issues, such as logical bugs in isolating abnormal nodes and mounting remote storage.
Other defects can occur in modules like job scheduling (\eg abnormal preemption) and platform configurations (\eg network settings).
These platform failures are highly severe and prioritized, requiring SREs to spend a significant amount of time resolving them promptly.

\finding{Although framework and platform faults cause relatively fewer LLM training failures, their diagnosis and mitigation are more challenging. Thus, the reliability of the LLM training frameworks and platforms deserves attention.}

\vspace{-5pt}
\subsection{RQ3: Data Sources of Failure Diagnosis}
\label{sec:RQ3}

Troubleshooting LLM training failures is challenging due to the complexity and scale of components, stages, and resources involved.
Based on the analysis of diagnostic procedures documented in \failurenum failure reports, \textbf{the average time to diagnose LLM training failures is 34.7 hours, with approximately 41.9\% of failures requiring more than 24 hours for diagnosis.} This highlights the time-consuming and labor-intensive nature of the diagnostic process.
To support failure diagnosis and better understand runtime behavior, LLM platforms, including \platform, are typically equipped with a variety of monitors that collect runtime information, such as training logs, performance metrics, and network traffic data. In this research question, we investigate the monitoring data sources typically used in the diagnosis process to better understand and improve failure diagnosis in LLM training platforms.

Similar to traditional software, training logs from different components capture detailed runtime information about the training procedure, offering valuable insights for users and engineers to understand the system's status~\cite{yuan2012conservative,yuan2020distributed,he2021survey}. Consequently, logs are a top priority for diagnosing LLM training failures in \platform in practice.
We meticulously reviewed the diagnosis processes recorded in all \failurenum failure reports and categorized each training failure into three diagnostic types: \textbf{(1) Log-only diagnosable}: Only training logs are involved in the diagnosis process. \textbf{(2) Non-log diagnosable}: Training logs do not provide useful clues for diagnosis. \textbf{(3) Hybrid diagnosable}: Both training logs and other monitoring data (\eg performance metrics) are jointly used for diagnosis.

The distribution of failure diagnostic types is shown in Tab.~\ref{tab:diagnosis_value}.
Notably, 53.9\% of LLM training failures can be diagnosed using training logs alone, without additional monitoring data.
This is because these logs, which include information from the training process, framework, hardware, and platform, provide comprehensive runtime details essential for diagnosing various types of faults in many cases.
For instance, if there is an error log \fixedwidth{``The ranktable or rank is invalid,Reason:[\%s].’’}~\cite{ranktable}, SREs can immediately notice the issues with the parallelism rank configurations and manually inspect the configuration files to determine the root causes.

However, there are also 10.1\% of training failures where training logs do not aid in the diagnosis.
In these cases, the logs either lack the necessary failure-indicating information or do not reflect the failure at all.
Consequently, SREs cannot rely on the LLM training logs to localize the faults.
Additionally, 36.0\% of training failures fall into the category of hybrid diagnosable failures.
In diagnosing these failures, SREs typically begin by examining the training logs to identify potential faulty components.
If the logs do not provide sufficient information to pinpoint the exact faulty components and root causes, SREs must then investigate other monitoring data to aid in the diagnosis.
These additional monitoring data typically include metrics of operator delay, GPU utilization rate, network packet loss, disk I/O rate and node heartbeats, which can help identify issues that are not explicitly reflected in the training logs.

In conclusion, training logs are the most crucial data source for diagnosing LLM training failures, with approximately 90\% of such failures requiring the information contained within these logs for diagnosis. However, the volume of training logs can be substantial, as each distributed process rank generates logs independently, and training durations are often extensive. \textbf{The average size of training logs for the failures we studied is 16.92GB.} Consequently, manually checking and identifying the failure-indicating logs within this volume of data is time-consuming and labor-intensive.
Moreover, 46.1\% of training failures cannot be diagnosed solely through training logs and require supplementary system monitoring data. This finding highlights the importance of developing comprehensive monitoring systems for LLM training platforms to improve the efficiency of failure diagnosis and facilitate rapid failure recovery.

\begin{table}[]
\renewcommand{\arraystretch}{1.2}
\centering
\caption{LLM Training Failures Across Diagnosis Types}
\vspace{-10pt}
\begin{tabular}{cccc}
  \toprule
  Diagnosis Type & Log-only & Non-log & Hybrid \\
  \hline
  Percentage  & 53.9\% & 10.1\% &36.0\% \\
  \bottomrule
\end{tabular}
\label{tab:diagnosis_value}
\vspace{-15pt}
\end{table}

\finding{Training logs are invaluable for diagnosing most (89.9\%) LLM training failures, but their large volume underscores the need for advanced log diagnostic tools for identifying the failure-indicating logs.}

\section{Automation Opportunities}

Our study results show that the automated identification of failure-indicating logs from large-scale training logs is crucial to enhancing the efficiency of failure diagnosis.
Therefore, in this section, we explore the automated opportunities for diagnosing LLM training failures based on training logs.

\subsection{Limitation of Existing Approaches}
\label{sec:limitation}

Numerous studies~\cite{chandola2009anomaly,lin2016logcluster,lin2020fast, rosenberg2020spectrum,liu2023scalable,huang2024demystifying} have focused on detecting anomalous logs in software systems.
These methods leverage features such as logging level~\cite{xu2009largescale}, event frequency~\cite{lin2016logcluster}, and error semantic~\cite{le2022log} to distinguish anomalous logs.
The anomalous logs detected serve as a potential failure indicator for fault diagnosis.

We have applied existing log-based anomaly detectors to LLM training logs on \platform, but our SREs reported that these methods struggle to distinguish failure-indicating logs from unrelated ones.
This issue stems from inherent limitations of the characteristics utilized by these detectors, rendering them ineffective for LLM training logs.
To better understand these characteristics, we randomly sampled 100 failures and manually labeled the failure-indicating logs within their training logs according to the documented failure diagnosis procedure.
Then, we analyzed the logging level, event frequency, and error semantics of these training logs.

\begin{figure}[t]
  \centering
  \mbox{
     \vspace{-2pt}
     \subfigure[Different logging levels.\label{fig:RQ4-level}]
     {
        \includegraphics[width=0.43\columnwidth]{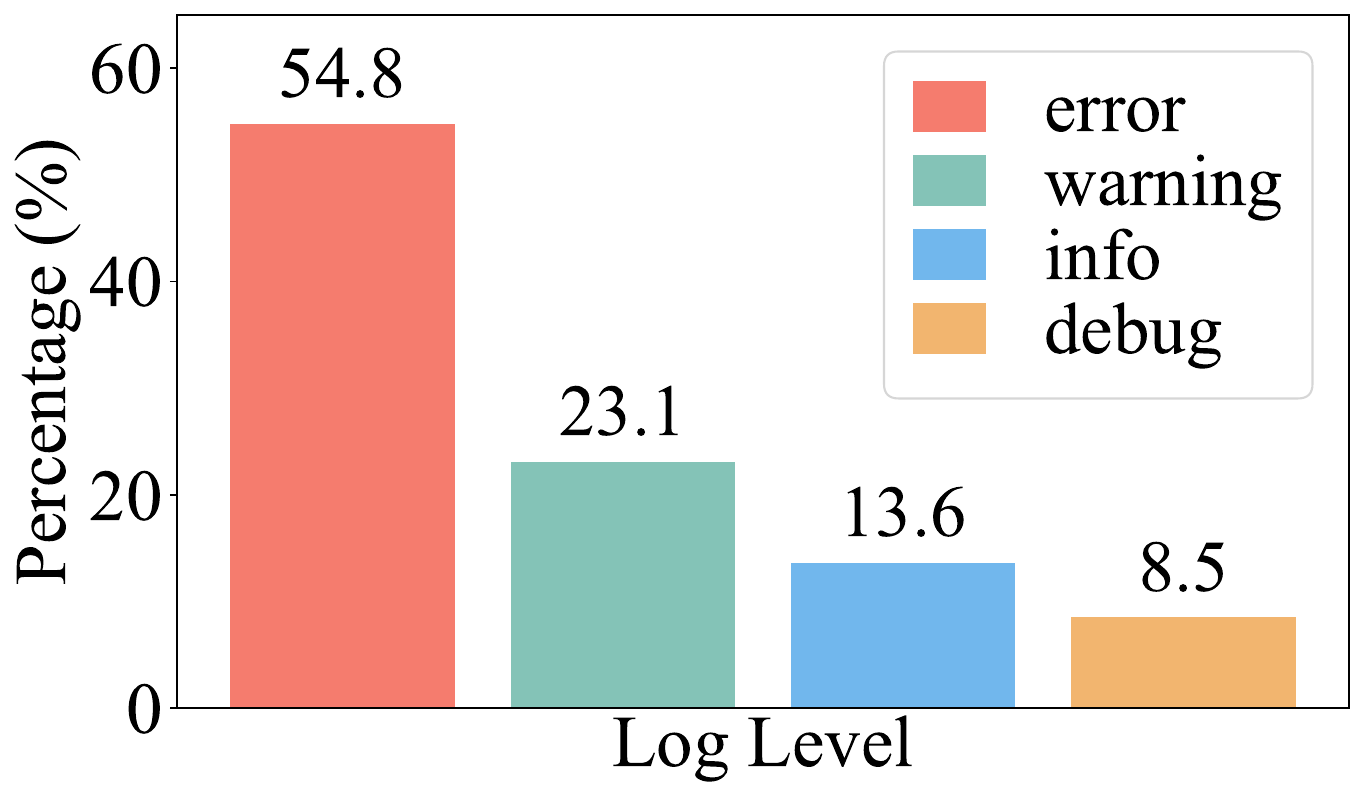}
     }
     \hspace{-6pt}\quad
     \vspace{-2pt}
     \subfigure[Different event frequencies.\label{fig:RQ4-frequency}]
     {
        \includegraphics[width=0.43\columnwidth]{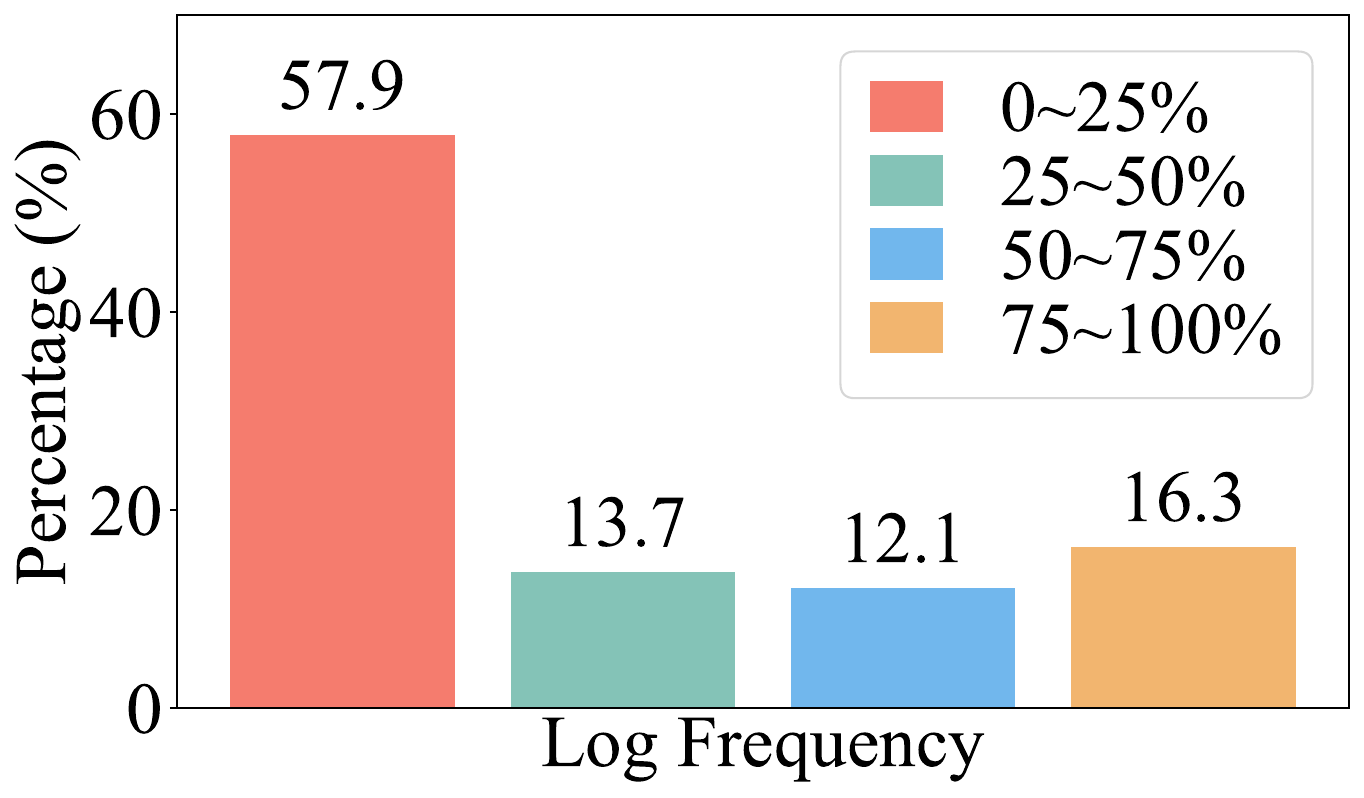}
     }
  }
    \vspace{-15pt}
  \caption{Distributions of LLM failure-indicating logs.}
    \vspace{-20pt}
\end{figure}

\noindent\textbf{Logging Level.}
Logging levels (\eg error, warning, info, and debug) indicate log importance.
Traditional log analysis methods~\cite{yuan2012conservative,bogatinovski2022leveraging} prioritize more serious logs, such as those at the error level.
We examined the distribution of failure-indicating logs across different levels in our sample dataset, as shown in Fig.~\ref{fig:RQ4-level}.
It is evident that about half (54.8\%) of these logs are at the error level.
The rest are spread across all log levels, including info (13.6\%) and debug (8.5\%), showing that logs at various levels can provide valuable insights for failure diagnosis.
Furthermore, we observed that many error-level logs are not related to training failures.
These discrepancies arise because log levels, which are determined by individual LLM training component, do not always reflect the overall severity and urgency in the training process.
For example, a log with failed checkpoint writing to remote storage might be logged in the error level by the checkpointing module, but if the fault-tolerance design allows for successful rewriting, this log does not impact the training process and thus is unrelated to failures.    

\noindent \textbf{Event Frequency.}
Event frequency is commonly used to detect anomalous logs, based on the intuition that infrequent log events are more likely to be anomalous~\cite{xu2009detecting,lin2016logcluster}.
However, this assumption does not hold for LLM training logs.
On the one hand, most infrequent logs are not failure-indicating logs.
Due to the numerous stages and steps in the LLM training procedure, many logs occur infrequently or even once during the entire process.
However, most of these infrequent logs are unrelated to diagnosing training failures.
On the other hand, failure-indicating logs are not necessarily infrequent, especially those during the training iteration phase.
We analyzed the event frequencies of failure-indicating logs in our sample dataset, as shown in Fig.~\ref{fig:RQ4-frequency}.
We categorized these logs into four groups based on their occurrence frequency, corresponding to the percentiles of lowest 0-25\%, 25\%-50\%, 50\%-75\%, and 75\%-100\%.
The results show that although more than half (57.9\%) of the failure-indicating logs fall within the lowest 25\% frequency, a notable portion still occur frequently, \eg 16.3\% of these logs are within the highest 25\% frequency.
Therefore, relying solely on frequency to identify failure-indicating logs is infeasible.

\noindent \textbf{Error Semantic.}
Recent work~\cite{le2022log,li2020SwissLog} has leveraged deep learning models, such as language models, to detect anomalies by analyzing the semantics of logs.
Logs with error semantics are flagged as anomalous and potential failure indicators.
However, these methods fall short in detecting failure-indicating logs in LLM training logs.
Firstly, not all logs with error semantics indicate failures.
Logs with error semantics from specific components or stages may not affect the training process or lead to failures.
We have observed that even in some successful training jobs, there are logs with errors in building wheels or recording hardware status, which are unrelated to failures.
Secondly, not all failure-indicating logs exhibit error semantics.
Some training failures manifest through abnormal behaviors rather than explicit error messages, making them undetectable by current semantic-based methods.

\finding{Existing log-based anomaly detectors struggle to effectively identify failure-indicating logs, as traditional anomalous log characteristics (\ie level, frequency and semantic) are not suitable for LLM training log scenarios. More effective approaches tailored for LLM training logs are needed.}

\subsection{Distinct Patterns of LLM Training Logs}
\label{sec:method_motivation}

Despite the inherent limitations of existing log analysis methods, we have observed three distinct patterns that can be used to automatically pinpoint failure-indicating logs.

\begin{itemize}[leftmargin=0pt, topsep=0pt, label={}]
    \item \textbf{Cross-job Pattern.}
    In practice, each failed training job is usually associated with a series of successful jobs with identical settings (\eg models and frameworks).
    For example, users typically validate configurations on a small scale of nodes before scaling up.
    Therefore, when analyzing a failed training job's logs, it is useful to review the logs of historical successful jobs with the same settings.
    As discussed in Sec.~\ref{sec:limitation}, even normal training jobs can produce numerous noisy error logs.
    Comparing logs from successful and failed jobs can help filter out unrelated noise and identify cross-job patterns.
    \item \textbf{Spatial Pattern.}
    Different from traditional software, the workflow of nodes in LLM training systems is highly synchronized and nearly identical.
    Consequently, the distributed log sequences generated by different nodes are very similar.
    As noted in Sec.~\ref{sec:RQ2}, local faults such as hardware faults cause a significant proportion of training failures.
    In such cases, logs from the faulty node can exhibit different patterns compared to others.
    Therefore, this spatial pattern in LLM training logs allows for comparison across nodes, enabling identification of differential logs that may indicate potential failures.
    \item \textbf{Temporal Pattern.}
    The LLM training procedure consists of multiple stages, each with distinct log characteristics.
    These features can help identify the stage where a failure occurred and filter out unrelated logs.
    For instance, if the iterative training stage has successfully started, error logs from the data and model loading stage are likely irrelevant.
    Furthermore, as discussed in Sec.~\ref{sec:RQ1}, most LLM training failures occur during the iterative training stage, where the workflow of all nodes is periodic and identical for each iteration.
    This temporal pattern can be used to compare logs across different iterations to automatically identify failure-indicating logs.
\end{itemize}

\finding{LLM training logs display special cross-job, spatial and temporal patterns, which can be leveraged to automatically identify failure-indicating logs.}
\vspace{-10pt}
\section{Our Framework: \nm}
\label{sec:framework}

\begin{figure*}[t]
    \centering
    \includegraphics[width=1.88\columnwidth]{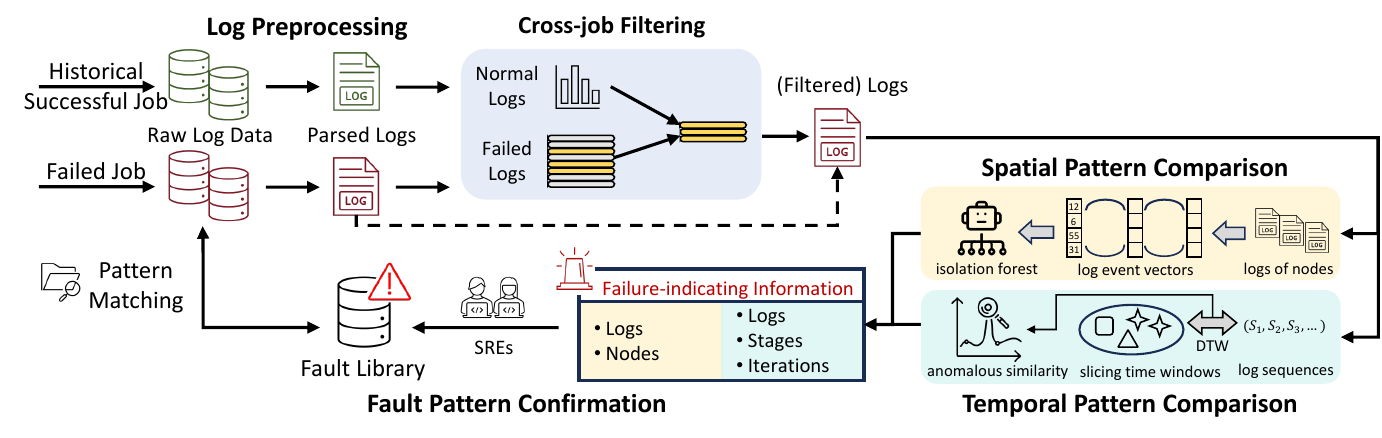}
    \vspace{-10pt}
    \caption{The overall framework of \nm.}
    \label{fig:method_framework}
    \vspace{-15pt}
\end{figure*}

Building on these observations, we propose \nm, a \underline{L}og-based \underline{L}arge-scale \underline{LL}M training failure diagnosis framework designed to automatically extract failure-indicating information from extensive training logs and improve diagnostic efficiency.
The overall framework is illustrated in Fig.~\ref{fig:method_framework}, containing four main phases. 

\vspace{-5pt}
\subsection{Log Preprocessing}
The raw logs of a training job are substantial (\eg tens of gigabytes) and unstructured, making them unsuitable for analysis.
To address this issue, we first preprocess the unstructured training logs through log parsing, transforming them into structured log events.
Specifically, log parsing aims to extract the constant parts of logs as templates and the variable parts as parameters, which has been widely studied~\cite{jiang2024large,jiang2024lilac,huang2024lunar}.
This structured information facilitates the identification of identical log events with varying parameters, enabling subsequent automated log analysis.
In this work, we adopt the widely-used and most efficient log parser, Drain~\cite{he2017drain}, for log preprocessing.
Following this parsing process, all raw logs are converted into sequential log events for further analysis.

\vspace{-5pt}
\subsection{Cross-job Filtering}

As outlined in Sec.~\ref{sec:method_motivation}, it is common for users to submit LLM training jobs with large-scale training nodes after successfully experimenting with smaller-scale ones~\cite{zhang2020empirical}.
This practice generates a history of successful jobs that share similar settings (\eg trained models and environments) with the failed job.
Intuitively, log events that are present in both successful and failed jobs are unlikely to indicate the failure root causes.
Therefore, these log events can be filtered out to reduce noise and enhance the analysis efficiency.

To achieve this filtering process, we first gather all log events from the parsed logs of the historical successful jobs to construct a normal log event pool, denoted as $\mathcal{N} = \{e_{n1}, e_{n2}, \cdots\}$.
Next, we examine the parsed logs of the failed job and remove all log events that are frequently present in $\mathcal{N}$ while preserving the original chronological order of the logs.
This step typically results in a significant reduction in the volume of logs compared to the original unfiltered logs, which is beneficial for subsequent analysis as it reduces the amount of extraneous information.
In cases where historical successful jobs are not available, we proceed by directly using the parsed logs of the failed training job for further processing.

\vspace{-5pt}
\subsection{Spatial and Temporal Pattern Comparison}
Inspired by the spatial and temporal pattern discussed in Sec.~\ref{sec:method_motivation}, this phase aims to identify failure-indicating information, such as log events, nodes, stages and iterations, by analyzing the spatial distribution and temporal sequence of training logs.
These types of interpretable information can be easily understood by engineers, facilitating more effective troubleshooting and in-depth analysis.

\subsubsection{Spatial pattern comparison}
Since the workloads distributed across nodes during the LLM training are nearly identical, it is expected that the training logs from these nodes will display similar patterns.
Therefore, examining the divergences in these patterns can reveal potential failure-indicating logs and suspicious nodes.
To facilitate this analysis, we first transform the parsed log events of each node into log event vectors, following previous studies~\cite{xu2009detecting, lou2010mining}.
Specifically, the log event vector for each node is denoted as $V = [c_1, c_2, \cdots, c_n]$, where $n$ represents the total number of distinct log events in the training logs and $c_i$ denotes the count of events of the $i$ -th log event.
This representation effectively captures both the occurrence and frequency information of training log events, providing an overview of training logs for each node.

After vectorization, we employ the Isolation Forest (iForest)~\cite{liu2008isolation} algorithm to detect deviant log event vectors across these nodes.
The iForest algorithm constructs a collection of isolation trees that isolate anomalies by randomly selecting a feature and then randomly selecting a split value between the maximum and minimum values of the selected feature.
The depth of a sample, averaged across the forest, serves as its anomaly score, with samples exhibiting noticeably shorter average path lengths being more likely to be anomalies.
We select the iForest algorithm for two primary reasons:
(1) It is an unsupervised method that does not require labeled data, making it highly adaptable to various training jobs.
(2) It provides interpretability by indicating the anomalous degree of each log event vector, which allows us to identify not only anomalous nodes but also specific log events contributing to the anomalies.
Finally, our spatial pattern comparison module utilizes the results from the iForest algorithm to pinpoint suspicious nodes and potential failure-indicating logs, which are then used for further troubleshooting.

\subsubsection{Temporal pattern comparison}
As discussed in Sec.~\ref{sec:background-training_platform}, LLM training can be divided into multiple stages.
These stages are typically recorded in the training logs, as LLM training frameworks often generate logs that denote the commencement and conclusion of each stage like data loading, model initialization.
Consequently, in this module, \nm first employs predefined rules to categorize identified logs in earlier phases into different stages.
This stage information helps SREs in broadly determining the stage at which the root causes may occur, thereby facilitating more efficient diagnosis.

Furthermore, the findings in Sec.~\ref{sec:RQ1} reveal that most LLM training failures occur during the iterative training phase, where each iteration involves executing a similar workload.
Thus, log event sequences across iterations are expected to follow consistent patterns.
Consequently, a log sequence that significantly deviates from those of preceding iterations could indicate a problematic iteration where a fault may occur.
To identify such failure-indicating iterations, we initially transform the parsed log events from each iteration into sequences of events, denoted as $S_i = [e_1, e_2, \dots, e_k]$.
We then choose the dynamic time warping (DTW) distance~\cite{muller2007dynamic}, which can dynamically align similar patterns in log sequences, to assess the similarity between log sequences from different iterations.
Subsequently, we select the slicing time window with ten iterations to calculate the average similarity scores between the log sequences of the current iteration and those of its antecedent iterations within this window.
Upon computing the similarity scores for all iterations, we apply the three-sigma rule~\cite{pukelsheim1994three} to detect any anomalous scores that are significantly lower than the average.
If such anomalies are detected, we identify and recommend these suspicious iterations and the logs within them as potentially failure-indicating information.

\vspace{-5pt}
\subsection{Fault Pattern Confirmation}

Following the aforementioned three phases of log analysis, \nm can automatically extract potential failure-indicating information, including suspicious logs, nodes, stages, and iterations from substantial training logs.
This automation circumvents the need for manual examination of extensive raw training logs, thereby significantly enhancing the diagnostic efficiency of training failures.
Additionally, the recommended information offers valuable insight for SREs, enabling them to swiftly understand the behavior and status of failures and accurately pinpoint their root causes.
After completing the diagnosis procedure, SREs summarize the log-based fault pattern of the failure and archive it in the historical fault library.
These confirmed fault patterns can be later used to directly match the training logs of new failed jobs, thereby improving the efficiency of handling similar failures in the future.

\section{Evaluation}
\label{sec:eval}

\subsection{Experiment Designs}

\noindent
\textbf{Evaluation Objective.}
We evaluate the effectiveness of \nm by answering the following two research questions (RQs):
\begin{itemize}[leftmargin=0pt, topsep=0pt, label={}]
    \item \textbf{RQ4:} How effective is \nm in identifying failure-indicating logs? 
    \item \textbf{RQ5:} How effective is \nm in locating faulty nodes?
\end{itemize}

\noindent
\textbf{Dataset.}
Our dataset comprises log files from 100 randomly sampled failed distributed LLM training jobs in our study dataset in Sec.~\ref{sec:study}, each averaging 632 accelerators.
Each accelerator corresponds to an individual process rank, thus generating a distinct log file.
As a result, the average log volume of each failed job is 12.3 GB. 
We manually labeled the failure-indicating log events for each case based on the recorded diagnostic procedure outlined in the corresponding failure report.
Specifically, within these cases, 42 were caused by hardware faults related to specific network devices, accelerators, and nodes.
For these cases, we marked the machines directly associated with the faulty component as the faulty nodes.

\noindent
\textbf{Baselines.}
(1) In RQ4, we compare \nm with three state-of-the-art, open-source log anomaly detection methods: LogAnomaly~\cite{meng2019loganomaly}, LogRobust~\cite{zhang2019robust}, and NeuralLog~\cite{chen2021neurallog}.
These baseline methods rely on labeled log data for training to achieve high accuracy.
Accordingly, we train the models using log data from 30\% of failed LLM training jobs and test their performance on the remaining 70\%.
(2) In RQ5, since no existing log-based methods are designed for faulty node localization, we compare \nm with two simple baselines, \textit{Error\_time} and \textit{Error\_count}, based on the practical intuition that nodes with earlier or more frequent error logs are more likely to be faulty.
In detail, \textit{Error\_time} ranks nodes by the time of their first error log, while \textit{Error\_count} ranks nodes by the total number of error logs.

\noindent
\textbf{Metrics.}
For RQ4, we compare the labeled failure-indicating logs with the detected anomalous logs, using precision, recall, and F1-score as evaluation metrics, following previous work~\cite{meng2019loganomaly,chen2021neurallog}.
For RQ5, since \nm can rank the detected suspicious nodes based on the anomaly degree, we use top-k accuracy to assess the results, is calculated by the proportion of the labeled faulty nodes that are included in the top-k candidates.

\subsection{Experiments Results}

\begin{figure}[t]
    \centering
    \includegraphics[width=0.95\columnwidth]{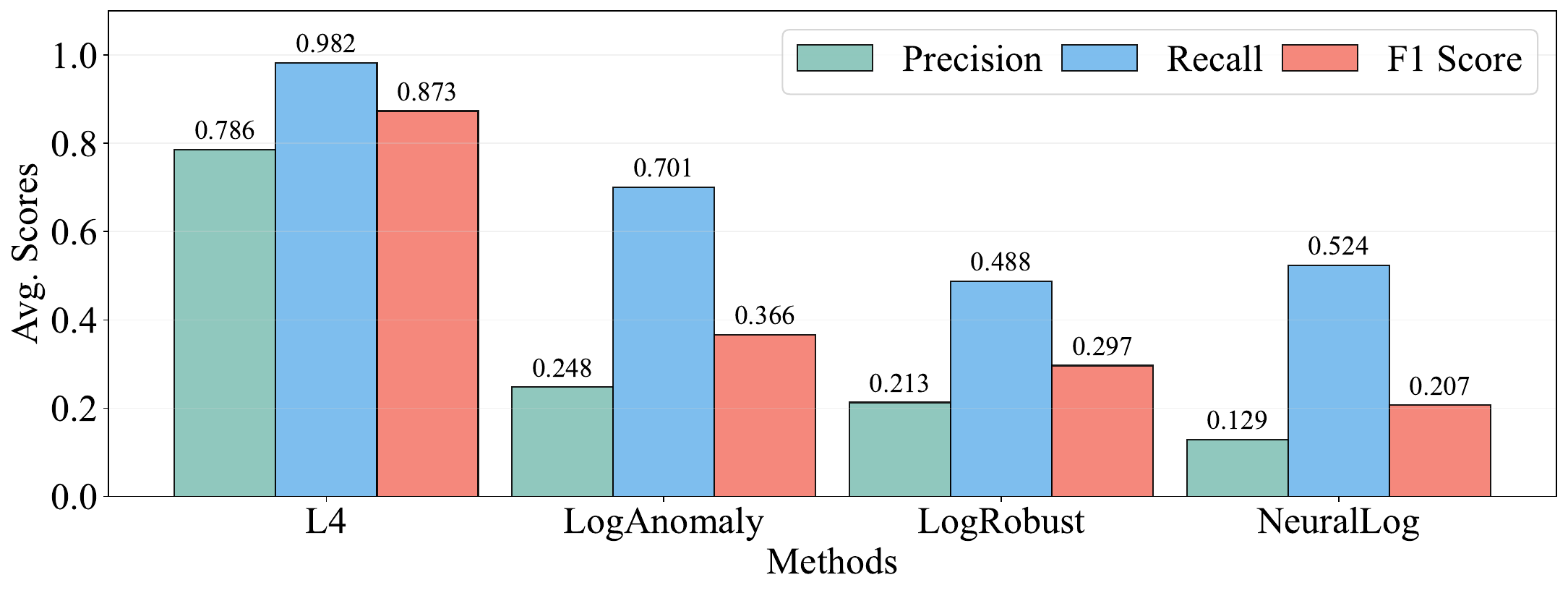}
    \caption{Effectivenss for identifying failure-indicating logs.}
    \label{fig:RQ4}
\end{figure}

\noindent
\textbf{RQ4: The effectiveness of identifying failure-indicating logs.}
We first conduct experiments to measure the capabilities of \nm and other cutting-edge log-based anomaly detection methods in LLM training failure scenarios.
Specifically, precision, recall, and F1-score are calculated by contrasting the ground-truth failure-indicating log events with the detected anomalous log events.

The experimental results are presented in Fig.~\ref{fig:RQ4}.
It is noteworthy that the existing log-based anomaly detectors exhibit poor performance when analyzing LLM training logs.
Evidently, the F1-scores of all baselines are considerably lower, ranging from 0.207 to 0.366.
This is primarily due to the low precision of these methods, as they fail to account for the specific patterns within distributed training logs, thereby resulting in false positive reports of anomaly logs.
Furthermore, their recall scores are also inferior to that of \nm, and there are two main reasons for this.
First, as investigated in Sec.\ref{sec:limitation}, many failure-indicating logs do not conform to the traditional anomalous characteristics (\eg low frequency) utilized by these methods.
Second, these methods rely heavily on labeled training data, and when dealing with new log data (\eg LLM training jobs with different frameworks), their accuracies are restricted.

In contrast, \nm consistently surpasses other baselines across all metrics.
In particular, \nm attains a high average recall of 0.982, demonstrating its robust ability to precisely identify failure-indicating logs without omission.
Although \nm’s precision (0.786) is marginally lower than its recall due to the existence of noisy logs, it still remains substantially higher than that of the other baselines.
Given that the detected results are intended for further examination by SREs, recall is prioritized over precision.
The high F1-score of \nm, exceeding baseline methods by over 0.507, further validates its superior balanced performance.

\noindent
\textbf{RQ5: The effectiveness of locating faulty nodes.}
As discussed in Sec.~\ref{sec:RQ2}, hardware fault is the most common root cause of LLM training failures.
\nm can unsupervisedly rank potentially faulty nodes based on their anomaly degrees through spatial pattern comparison.
In this RQ, we evaluate the accuracy of \nm and two baselines in finding faulty nodes in cases of hardware-related training failures.

\begin{figure}[t]
    \centering
    \includegraphics[width=0.95\columnwidth]{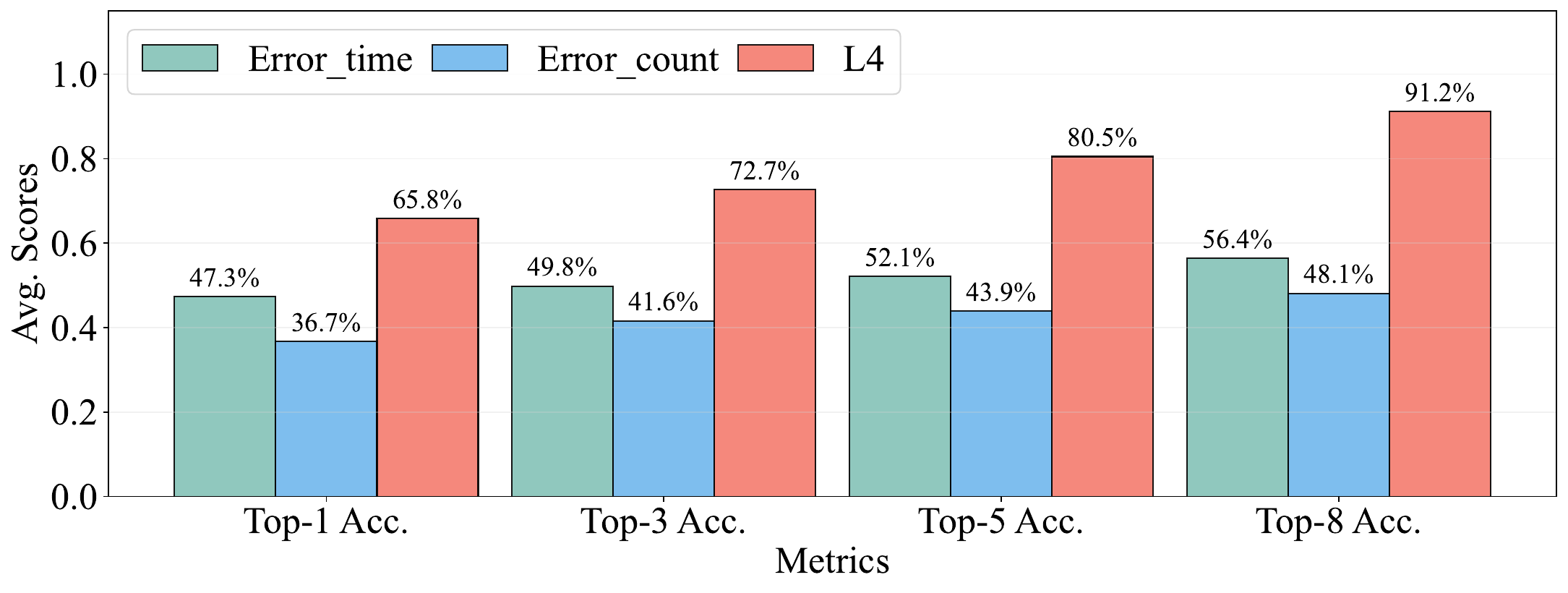}
    \caption{Effectivenss for locating faulty nodes.}
    \label{fig:RQ5}
\end{figure}

As shown in Fig.~\ref{fig:RQ5}, \nm significantly outperforms two baselines across all metrics, with improvements ranging from 18.5\% to 43.1\%.
For example, the top-1 accuracy of \nm in localizing faulty nodes is 65.8\%, compared to 47.3\% and 36.7\% for \textit{Error\_time} and \textit{Error\_count}, respectively.
This demonstrates that relying solely on error log time and count is insufficient for identifying faulty nodes.
Furthermore, \nm’s accuracy increases with the number of candidate nodes, reaching 91.2\% when considering the top 8 candidates.
These results indicate that \nm effectively detects anomalous log patterns in faulty nodes under hardware-related failure scenarios by leveraging spatial patterns.
It is also worth noting that not all faulty nodes will exhibit the most anomalous log patterns.
In some cases, nodes associated with the faulty node (\eg those in the same communication area) may also produce outlier log patterns.
Consequently, \nm recommends multiple top-ranked anomalous nodes (the top 8 by default) that exceed an anomalous degree threshold for further investigation by SREs.
In large-scale LLM training scenarios involving thousands of nodes, such recommendations greatly enhance diagnostic efficiency by narrowing the scope of investigation.

\section{Deployment Experience}

\nm has been successfully applied in the failure management system of \platform to analyze the training logs of failed LLM training jobs since June 2024.
Once a failure report is reported, \nm first attempts to match its training logs with historically confirmed fault patterns in the fault library.
If the match is successful, the recorded root cause and fix solutions are retrieved and recommended to SREs.
Otherwise, \nm automatically analyzes the extensive training logs by identifying failure-indicating log events, nodes, stages and iterations, which are then recommended to SREs for further understanding and diagnosis.
This procedure significantly reduces the manual effort required to investigate substantial raw training logs.
In the following, we depict two real-world cases illustrating the effectiveness of \nm in diagnosing LLM training failures.

\noindent
\textbf{Case 1: Fine-grained Localization of Hardware Fault.}
As studied in Sec.~\ref{sec:RQ2}, hardware faults are the most common cause of LLM training failures.
Localizing these faults to specific hardware (\eg nodes, accelerators, and network links) was previously labor-intensive and time-consuming due to the large scale of training resources.
However, with \nm, SREs can promptly pinpoint the faulty hardware causing training failures.
For instance, an LLM training job involving 1024 nodes with 4096 accelerators on \platform failed after 20 hours of training, producing 71 GBs of latest training logs.
\nm was employed to analyze these substantial logs and it identified eight nodes with log event patterns that diverged from those of the other nodes through spatial pattern comparison.
Consequently, \nm recommended these suspicious nodes and log events such as \fixedwidth{`ROCE(,hccp\_service.bin):error cqe status.'}.
Based on this information, SREs inspected the indicated nodes and logs, uncovering a hardware fault in an accelerator on one of these nodes.
This fault also caused the nodes directly communicating with the faulty node to generate relevant error logs.
Following this discovery, the SREs isolated and replaced the faulty node, restored the training procedure, and subsequently summarized this fault pattern into the fault library for diagnosing similar failures in the future.

\noindent
\textbf{Case 2: Issue Identification during Iterative Training.}
According to Finding 1, most LLM training failures occur during training iterations.
\nm can effectively identify the failure-indicating iteration to aid in diagnosis.
For example, in \platform, an LLM training job hanged after around two thousand epochs.
\nm was applied to analyze the training logs, and through temporal pattern comparison, \nm identified an anomalous log event sequence in one iteration.
This sequence included additional logs, such as \fixedwidth{`notify wait from rank\_<*> timeout'}, which were present in this iteration but absent in preceding ones.
This failure-indicating information inspired SREs to suspect a network fault, prompting them to check the network packet loss rate for this ranked node.
They found a spike during the time frame of the anomalous iteration, confirming intermittent network faults in the communication links, which caused timeouts and stalled the training.
Consequently, SREs isolated and repaired the faulty network links for failure recovery, incorporating this fault pattern into the fault library for future failure diagnosis.

These two typical cases demonstrate that \nm can effectively contribute to diagnosing the majority of LLM training failures.
However, there remains a small portion of cases where \nm cannot directly pinpoint the exact root cause.
For instance, in rare cases involving logic bugs within LLM training frameworks, \nm is unable to locate the specific faulty code.
Nonetheless, in such situations, \nm can still provide valuable diagnostic information (\eg the failed stage) and filter out failure-unrelated noisy logs.
This process reduces the manual effort required to analyze training logs and enhances the diagnostic efficiency of LLM training failures for SREs.

\section{Discussion}

\subsection{Generalizability}
\label{sec:generalizability}

Our study investigates \failurenum diverse LLM training job failures on the \platform. However, we believe that the studied LLM failures are common and representative, and that the findings can be generalized to other LLM platforms.

On the one hand, \platform employs widely adopted technologies and shares architectural similarities with other leading platforms~\cite{jeon2019analysis,jiang2024megascale,kokolis2024revisiting}, which utilize similar job management mechanisms.
Besides, most training jobs on \platform involve diverse open-source and commonly used models (\eg LLaMA~\cite{dubey2024llama} and Vicuna~\cite{chiang2023vicuna} series), frameworks (\eg PyTorch~\cite{pytorch} and Transformer~\cite{transformer}) and hardware, underscoring the commonality of our analysis.

On the other hand, our study avoids drawing conclusions that are overly specific or narrowly applicable to \platform.
In addition, reports from industry practitioners have identified similar issues in large-scale LLM training systems.
For instance, \citet{kokolis2024revisiting} classify LLM training failures into domains such as user programs, system software, and hardware infrastructure, aligning closely with our observations.
Similarly, \citet{hu2024characterization} and \citet{jiang2024megascale} highlight the diagnostic challenges of LLM training failures and underscore the critical role of training logs in the diagnosis process.
Our study provides a more comprehensive analysis of failure root causes and diagnostic procedures, along with the discussion of automation opportunities, offering actionable insights for other LLM platforms to improve system reliability.

Regarding our proposed framework, \nm, its applicability extends beyond \platform and can enhance failure diagnosis efficiency across different LLM platforms.
The three key log patterns leveraged by \nm—cross-job, spatial, and temporal—are not specific to \platform and are broadly applicable to various LLM training scenarios.
The successful deployment of \nm on a large number of LLM training jobs involving diverse models, frameworks, and hardware configurations further demonstrates its generalizability.

\subsection{Future Directions}

\noindent
\textbf{LLM Training Monitoring.}
According to Finding 5, 10.1\% of failures could not be diagnosed using the current platform's monitoring data and logs, indicating that existing monitoring systems for LLM training are insufficient and could be improved.
Future enhancements could incorporate additional signals like training data lineage and program profiling to offer a more holistic view of the training process.
Tracking the provenance of training data and analyzing the performance characteristics of training programs could enhance the platform's monitoring capabilities, enabling earlier detection and diagnosis of failures.
Additionally, our research highlighted issues with the quality of LLM training logs, characterized by excessive noise and low correlation between logging levels/semantics and failure relevance.
Future work can focus on providing recommendations on logging levels, locations, and contents to improve log standardization and informativeness in LLM training frameworks, ensuring that logged information is more closely tied to potential failures and facilitating efficient failure diagnosis~\cite{li2024go,li2023exploring}.

\noindent
\textbf{Multi-modal Failure Diagnosis.}
Our \nm is designed to extract failure-indicating information from extensive training logs, enhancing the efficiency of diagnosing LLM training failures.
However, analyzing logs solely cannot handle all failure types.
As shown in Sec.~\ref{sec:RQ3}, 36.0\% of failures require hybrid monitoring data for accurate diagnosis.
Future research could further integrate multi-modal monitoring data, similar to cloud system failure diagnosis~\cite{he2022perfsig,Yu2023Nezha,zhu2024hemirca}.
Combining diverse data modalities such as logs, performance metrics, and network profiling data can provide a comprehensive view of the training process, allowing for a more accurate and comprehensive failure diagnosis.

\section{Related Work}

\textbf{Model Training Failure Study.} Recently, a series of studies have focused on failures in deep learning (DL) platforms.
\cite{gao2023empirical} studied 360 quality issues of DL jobs, categorizing common symptoms, root causes and fixes, while \cite{zhang2020empirical} focused on program failures and reviewed current testing and debugging practices in DL platforms.
\cite{he2023understanding} presented an in-depth study on hardware faults of accelerators in DL systems.
However, as previous DL jobs differ significantly from LLM training jobs in model sizes, architectures, training resources, and software stacks~\cite{hu2024characterization}, new research on large-scale LLM training jobs is necessary to uncover their unique characteristics and associated failures.

The closest related work is by~\cite{hu2024characterization}, which examined a six-month LLM development workload and probed discrepancies between LLMs and prior DL workloads.
However, their analysis did not focus on training failures and diagnosis procedures.
In contrast, our work comprehensively examines failure symptoms and root causes, as well as automated opportunities for log-based failure diagnosis, offering valuable insights for both practitioners and researchers.

\textbf{Log-based Failure Diagnosis.} Logs are essential for ensuring software reliability and diagnosing issues by providing critical runtime information~\cite{amar2019mining, lin2020fast, yuan2020distributed}.
A significant body of research focuses on diagnosing by identifying anomalous logs for further investigation. 
For instance, \cite{xu2009detecting} pioneered the application of Principal Component Analysis (PCA) to detect system issues from console logs based on log event frequencies.
LogCluster~\cite{lin2016log} employs clustering to group similar logs, thereby identifying atypical log events.
SBLD~\cite{rosenberg2020spectrum} applied spectrum-based techniques to transform logs into events and identify key logs by evaluating event coverage.
Moreover, some supervised approaches, such as LogRobust\cite{zhang2019robust}, leverages log semantics for more effective analysis.
Recently, LLMs have been applied to analyze log semantics, enhancing the identification of anomalous logs with error semantics for diagnosis~\cite{liu2023scalable, pan2023raglog}.

However, as outlined in Sec.~\ref{sec:limitation}, these methods assume certain characteristics of anomalous logs that are inapplicable to the context of LLM training logs. 
In contrast, our \nm utilizes the distinct patterns in LLM training logs to accurately identify failure-indicating information.
Our study can also provide guidance and opportunities for future log-based LLM training failure diagnosis research.

\section{Conclusion}

In conclusion, this work provides the first comprehensive study of LLM training failures on \platform.
Our investigation reveals diverse root causes and diagnostic challenges of these failures, with existing log analysis methods being inadequate due to their reliance on traditional log characteristics.
We also identify three distinct patterns (\ie cross-job, spatial and temporal) within LLM training logs, which inspire the design of our \nm, a log-based large-scale LLM training failure diagnosis framework.
\nm significantly enhances diagnostic efficiency and accuracy by automatically extracting failure-indicating information from extensive training logs.
Our findings and \nm can offer valuable information for diagnosing LLM training failures and ensuring the reliability of LLM training systems.

\balance
\bibliographystyle{ACM-Reference-Format}
\bibliography{sample-base}

\end{document}